\documentclass[10pt]{article}
\usepackage[T2A]{fontenc}
\usepackage[cp1251]{inputenc}
\usepackage[english]{babel}
\usepackage{graphicx,caption,subcaption}
\usepackage{amsmath}
\usepackage{amssymb}
\usepackage{ulem}
\usepackage{makeidx}
\usepackage{fancyhdr}
\usepackage{chngcntr}
\usepackage{caption}
\usepackage{subcaption}

\begin{document}

\begin{center} \noindent {\textbf{\Large A homogenization theory for systems of
penetrable dielectric particles }}

\bigskip

{M.~Ya.~Sushko\footnote{ Corresponding author, e-mail:
mrs@onu.edu.ua}} and A.~V.~Dorosh

{\it Department of Theoretical Physics and Astronomy, Mechnikov
National University,

2 Dvoryanska St., Odesa 65026, Ukraine}

\end{center}

\begin{abstract}
A many-particle theory is presented for the effective quasistatic
permittivity of macroscopically homogeneous and isotropic systems
of inhomogeneous dielectric particles with different degrees of
penetrability.  The theory is based upon our original
compact-group approach, complemented by the Hashin-Shtrikman
variational principle. The governing equation is obtained by
summing up the statistical moments for the deviations of the local
permittivity in the system from the desired effective
permittivity. The latter is, in principle, recoverable from the
governing equation as a functional of the constituents’ volume
concentrations (expressed through statistical averages of certain
products of the particles' characteristic functions) and
permittivity profiles. Under the suggestion that the local
permittivity is determined by the shortest distance from the point
of interest to the nearest sphere, a complete analysis is carried
out for hard and fully penetrable spheres with
piecewise-continuous radial permittivities. The results are
contrasted  with other authors' analytical theories and simulation
data. This comparison validates our theory and also sheds light on
possible computational errors caused by the use of rectangular
lattices to simulate dispersions of spherical particles.
\end{abstract}

\vspace{2pc} \noindent{PACS numbers}: 02.70.--c, 42.25.Dd,
77.22.Ch, 82.70.--y

\vspace{2pc} \noindent{\it Keywords}: Permittivity, Dispersed
system, Penetrable particle, Compact group method,
Hashin-Shtrikman principle, Homogenization

%\maketitle

\section{Introduction}\label{c:introduction}

The question of how the statistical microstructure affects the
dielectric  properties of a particular substance, natural or
artificial, is of great importance to a wide range of physical,
chemical, biological, and interdisciplinary disciplines
\cite{Bohren1983,Bergman1992,Nan1993,Sihvola1999,Tsang2001,Torquato2002,Morgan2003,DelgadoEKP,Ohshima12}.
In particular, it lies at the heart of material science whose main
objectives include the development of composite materials with
desired dielectric parameters. The inverse problem of finding the
microstructure of a system from a known response of the latter to
an applied electric field is of no less significance.

\begin{figure}[!tbp]
  \centering
  \begin{minipage}[t]{0.45\textwidth}
    \includegraphics[width=\textwidth]{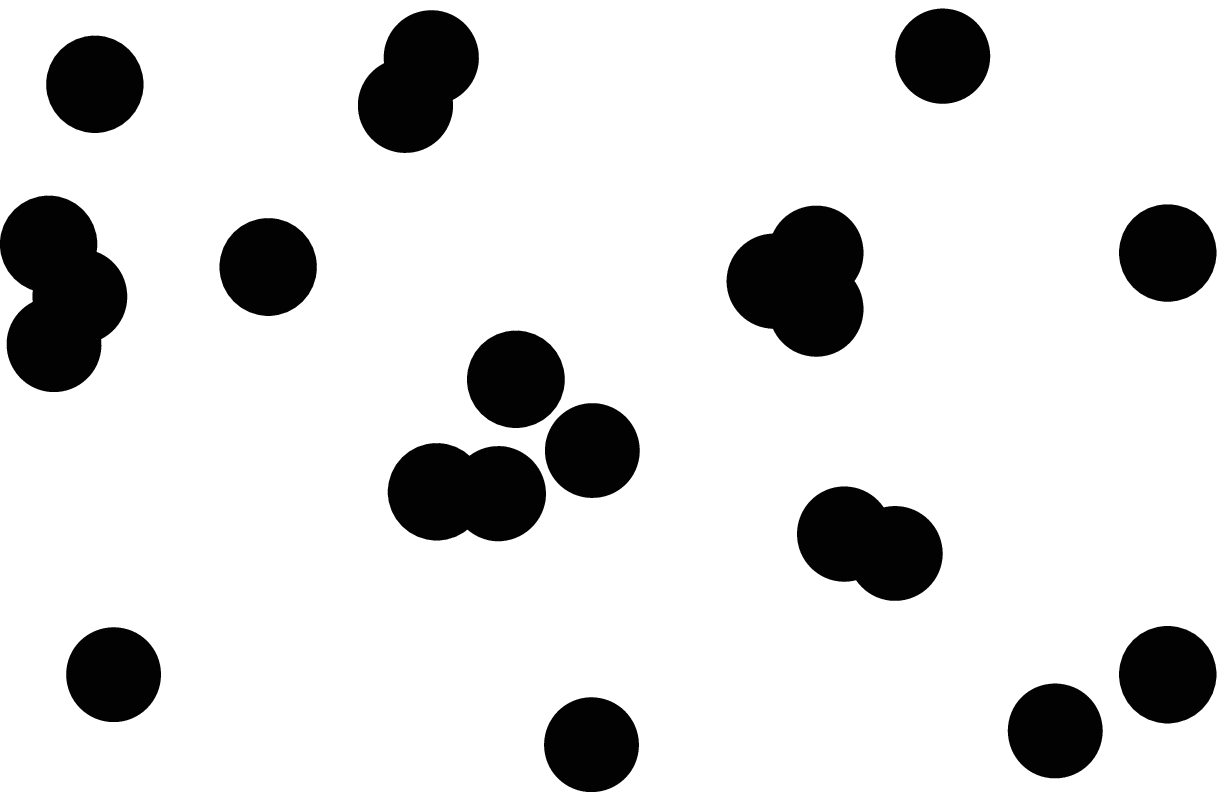}
    \caption{\label{fig:SoftBalls} {\normalsize A system of penetrable uniform spheres whose positions
are statistically independent and which are allowed to overlap to
form clusters with the same permittivity.}}
  \end{minipage}
  \hfill
  \begin{minipage}[t]{0.45\textwidth}
    \includegraphics[width=\textwidth]{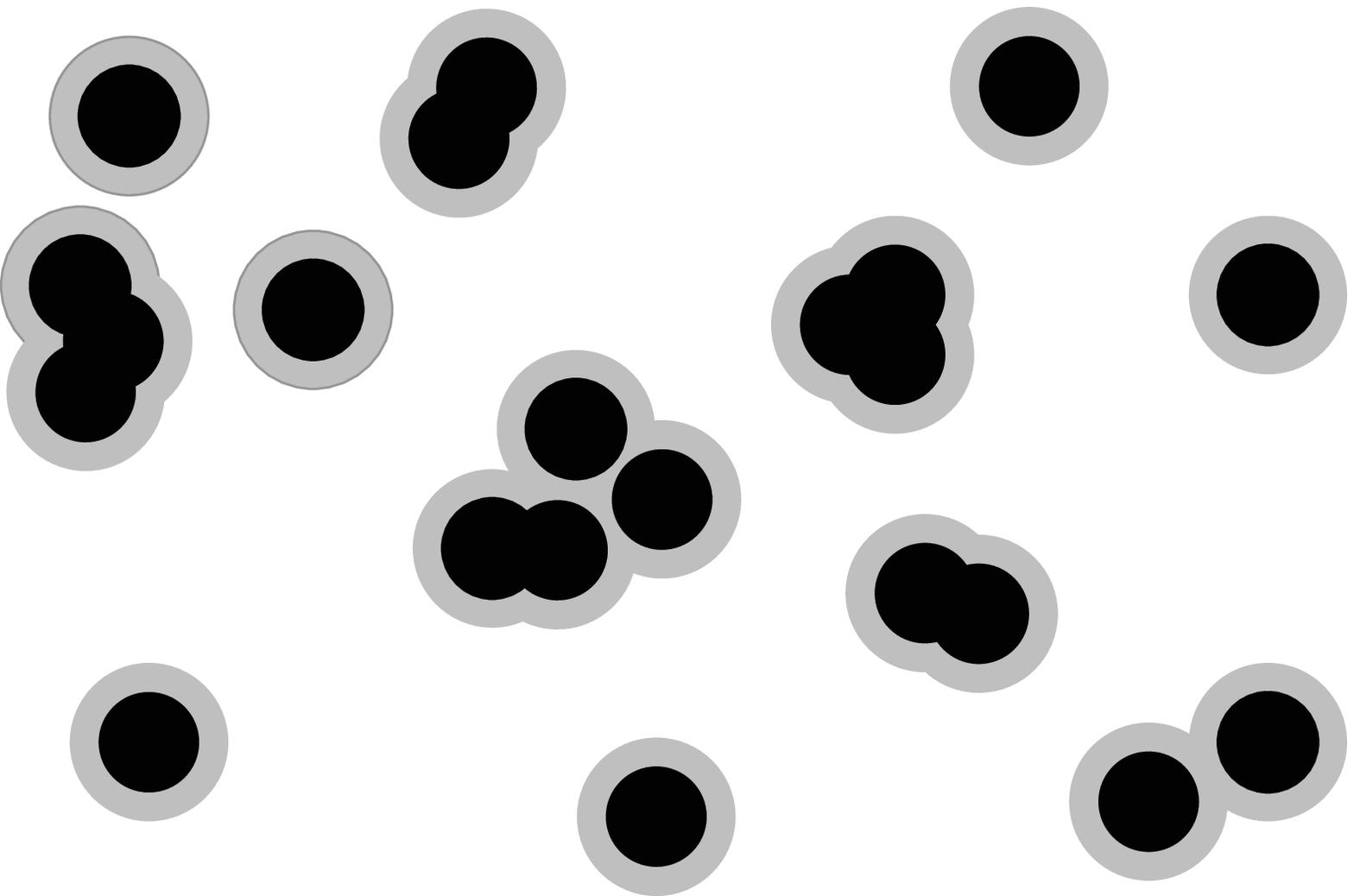}
    \caption{\label{fig:SoftLayeredBalls}{\normalsize A system of penetrable two-layer spheres. The local
permittivity is determined by the distance from the point of
interest to the nearest sphere's center. }}
  \end{minipage}
\end{figure}

In this report, we analyze the effective quasistatic permittivity,
$\varepsilon_{\rm eff}$, of those heterogeneous systems which can
be viewed as a result of dispersing dielectric particles with
different degrees of penetrability (understood as the capability
of particles to overlap one another) into uniform dielectric
matrices (see, for examples,  Figs.~\ref{fig:SoftBalls} and
\ref{fig:SoftLayeredBalls}). The permittivity $\varepsilon_{\rm
eff}$ characterizes the dielectric response of the system to a
probing electrical field with wavelength $\lambda \to \infty$ and
is defined as the permittivity of a homogeneous (homogenized)
medium producing the same dielectric response to the probing
filed. In fact, a pure substance with the properties of the
homogenized medium may not occur in nature.

The determination  of $\varepsilon_{\rm eff}$ is part of a general
many-particle problem called the homogenization of a heterogeneous
system. Solving it requires that both polarization and correlation
effects of higher orders be taken into account, and it is the
long-wave limit where the greatest progress has been achieved (see
\cite{Bohren1983,Bergman1992,Nan1993,Sihvola1999,Tsang2001,Torquato2002}
for a review). Yet even in this case, one is faced with serious
difficulties, especially when dealing with aperiodic or random
systems. Some facts behind this statement are as follows:

1. One class of homogenization theories, exemplified by
\cite{Tsang2001,Lax1952,Lamb1980,Tsang1980,Davis1985,Geigenmuller1986,Claro1991,Fu1993,Kuzmin2005,Mallet2005},
represents attempts to use the multiple scattering theory or, in
the quasistatic limit, the Green-function method for
boundary-value problems to derive multipolar corrections to the
classical Maxwell-Garnett \cite{Maxwell1873,Maxwell1904} and
Bruggeman \cite{Bruggeman1935,Landauer1952} mixing rules, which
were obtained long ago within one-particle and dipolar
approximations. The consideration is usually restricted to the
evaluation, using the Kirkwood superposition approximation and
model pair-distribution functions, of correlation effects in
groups of several hard particles. The precise role of the
discarded higher-order terms in the iterative series becomes
obscure as the filler concentration, or the dielectric filler-host
contrast, or both are increased.

2. Similar difficulties are typical of another class of
homogenization theories, such as
\cite{Torquato2002,Brown1965,Batchelor1972,Jeffrey1974,Felderhof1982,Felderhof1982a,Ramshaw1984,Torquato1984,Torquato1985b,Torquato1985},
based upon the perturbation expansion \cite{Brown1955} or cluster
expansion \cite{Finkelberg1963,Finkelberg1964a,Finkelberg1964}
technique. Their basic idea is to express the average polarization
as a formal operator (an ensemble-averaged quantity) acting on the
applied field and then use either technique to eliminate the
applied field in favor of the average electric field in the
system.   Besides the knowledge of many-particle distribution
functions, one requires the solutions to the electrostatic
boundary-value problems for clusters of particles to advance.
However, the exact solution is available only for two
nonoverlapping spheres \cite{Jeffery1912} and no such solutions
exist for overlapping spheres \cite{Torquato2002,Torquato1984}.
Even an approximate analysis  of the three-sphere boundary problem
is extremely difficult \cite{Cichocki1988,Ju1994,Buryachenko2001}.

3. One more class of homogenization theories, termed together as
the strong-property-fluctuation theory (SPFT) and represented by
\cite{Tsang2001,Ryzhov1965,Ryzhov1970,Tamoikin1971,Tsang1981,Zhuck1994,Michel1995,Mackay2000,Mackay2001},
is usually restricted to the use of the second-order truncation
(bilocal approximation) for the mass operator series in the
Dyson-type equation for the averaged electric field  (for a
discussion  of the third-order approximation, see
\cite{Mackay2001}), Gaussian statistics for the stochastic field,
model expressions for the two-point correlation function, and
several other approximations.

4. The analysis of the effects of the particles' inhomogeneity and
anisotropy on the overall response of the system is usually
reduced to the study of the response of a solitary particle to a
uniform field. However, for a system of overlapping particles, the
concepts of an individual particle and its polarizability become
ill-defined. It is therefore reasonable that the theory of
dielectric response of such a system  be elaborated in a way that
effectively incorporates many-particle effects, but avoids using
uncontrolled assumptions about those to the greatest extent
possible.

For diluted dispersions of penetrable spheres, such a theory was
developed in \cite{Torquato1984,Torquato1985b,Torquato1985}.
However, to our best knowledge, no consistent analytical approach
to $\varepsilon_{\rm eff}$ has been proposed so far for
concentrated systems of penetrable particles. Some simulation
results for freely overlapping spheres were obtained in
\cite{Pekonen1999,Karkainen2001}. A rather good interpolation of
those is given by differential mixing equation \cite{Jylha2007},
derived in a similar manner as the Maxwell Garnett mixing
equation, but by including an enhanced background permittivity
effect as the volume concentration of particles increases.

The main points of our approach to the problem and the structure
of the paper are as follows.

(1) We suggest that the dielectric response of a dispersion,
${\cal D}$, to be homogenized is equivalent to that of an
auxiliary model system, ${\cal S}$, made up by embedding all the
constituents of ${\cal D}$ into a uniform (perhaps imagined) host
medium, ${\cal M}$, with a permittivity $\varepsilon_{\rm f}$ to
be found.

(2) The effective permittivity of ${\cal S}$ (equal to
$\varepsilon_{\rm eff}$ of ${\cal D}$) is analyzed with the method
of compact groups
\cite{Sushko2007,Sushko2009CompGroups,Sushko2009AnisPart}. A
compact group is defined  as a macroscopic group of
inhomogeneities (${\cal D}$'s particles and/or regions) within
which all the distances between the inhomogeneities are much
smaller than the wavelength of probing radiation in the host
(${\cal M}$). Such groups are assumed to be large enough to
reproduce the properties of the entire ${\cal S}$. In the
long-wavelength limit, it is the multi-particle polarization and
correlation processes inside compact groups that give the leading
contributions to the iterative series for the electric field and
induction  in ${\cal S}$. On the hand, with respect to the probing
field compact groups are actually point-like.  Using the methods
of generalized function theory, their contributions can be singled
out without going into details of the above  processes. The basic
relations for ${\cal S}$ in terms of the compact-group method are
summarized in section \ref{c:compactgroups}.

(3) Finding $\varepsilon_{\rm f}$ is a separate problem. To solve
it, we additionally appeal \cite{Sushko2016,Sushko2017} to the
Hashin-Shtrikman variational principle \cite{Hashin1962} to obtain
$\varepsilon_{\rm f}=\varepsilon_{\rm eff}$. The details of the
solution are outlined in section \ref{c:compactgroups} as well.

(4) The permittivity distribution in  ${\cal S}$ is the sum of
$\varepsilon_{\rm f}$ and the contribution $\delta
\varepsilon({\bf r})$  due to compact groups of ${\cal D}$'s
constituents embedded in ${\cal M}$. The $\delta \varepsilon({\bf
r})$ is modeled in terms of the characteristic (indicator)
functions and permittivity profiles of individual constituents in
accordance with the rules defining a model. With $\varepsilon_{\rm
f}$ and $\delta \varepsilon({\bf r})$ known, finding
$\varepsilon_{\rm eff}$ reduces to a calculation and summation of
the statistical moments of $\delta \varepsilon({\bf r})$.

(5) For dispersions of uniform spheres with different degrees of
penetrability, this procedure gives Bruggeman-type equations for
$\varepsilon_{\rm eff}$ as a function of the permittivities and
effective volume concentrations of ${\cal D}$'s constituents.
These results and comparison of them with calculations
\cite{Jeffrey1973,Felderhof1982a,Torquato1984,Torquato1985},
differential mixing rule \cite{Jylha2007}, and extensive
simulation data \cite{Karkainen2001} are presented in section
\ref{c:formalismSpheres}. The discrepancies with the simulations
turn out to be too large to be accepted. We hypothesize that they
are caused by the ambiguity in determination of the permittivity
value between potential nodes near an arbitrarily-oriented
material interface. Setting this value to be different from the
permittivities of ${\cal D}$'s constituents is effectively
equivalent to an introduction of thin surface layers covering the
spheres.

(6) Generalizations of our theory to dispersions of fully
penetrable or hard isotropic spheres with piecewise-continuous
radial permittivity profiles $e=e(r)$ are suggested in section
\ref{c:inhomogeneusParticles}. In either case, $\varepsilon_{\rm
eff}$ is a functional determined by a certain integral relation.
Using the model of two-layer spheres and considering the thickness
and permittivity of the surface layer as fitting parameters,
simulation data \cite{Karkainen2001} can be reproduced within
their accuracy. Analytical results \cite{Sihvola1989} for hard
spheres  are recovered by our theory surprisingly well for all
continuous profiles $e(r)$ considered in there.

(7) The major results of the paper are summarized in section
\ref{c:conclusion}.

\section{Basics of the compact-group approach}\label{c:compactgroups}

Suppose that a fine dispersion ${\cal D}$ consists of a host, with
permittivity $\varepsilon_0$, and dispersed particles, with
permittivity $\varepsilon_1(\bf r)$. To find its effective
quasistatic permittivity, we assume that ${\cal D}$ is equivalent,
in its effective dielectric properties, to a macroscopically
homogeneous and isotropic system ${\cal S}$ prepared by embedding
${\cal D}$'s constituents into a certain (in general, imagined)
uniform medium ${\cal M}$, of permittivity $\varepsilon_{\rm f}$.
According to the compact-group approach
\cite{Sushko2007,Sushko2009CompGroups,Sushko2009AnisPart}, ${\cal
S}$ can be viewed as a set of compact groups of
 ${\cal D}$'s constituents  in ${\cal M}$. The local
permittivity in ${\cal S}$ is modeled as
\begin{equation}\label{localepsilon}
\varepsilon (\textbf{r})=\varepsilon_{\rm f}+\delta \varepsilon
(\textbf{r}),
\end{equation}
where $\delta \varepsilon (\textbf{r})$ is the deviation from
$\varepsilon_{\rm f}$ due to the presence  of a compact group at
point $\bf{r}$. The explicit form of $\delta \varepsilon
(\textbf{r})$ depends on the properties  of both the embedded
particles and the host.

The effective permittivity $\varepsilon_{\rm eff}$ of ${\cal S}$
(and that of ${\cal D}$) is determined  as the proportionality
coefficient in \cite{Landau1982}
\begin{equation}\label{D=eeffE}
\langle\textbf{D}(\textbf{r})\rangle=\langle\varepsilon
(\textbf{r})\textbf{E}(\textbf{r})\rangle =\varepsilon_{\rm
eff}\langle \textbf{E} (\textbf{r})\rangle,
\end{equation}
where $\textbf{D}(\textbf{r})$ and $\textbf{E}(\textbf{r})$ are
the local induction and electric field in ${\cal S}$,
respectively, and the angular brackets stand for statistical
averaging or averaging by integration over the volume $V$ of
${\cal S}$ (that is, $\langle{\bf E}({\bf r})\rangle = V^{-1}
\int_V {\bf E}({\bf r})\, d{\bf r}$, etc). The ergodic hypothesis
suggests  \cite{Landau1982,Torquato2002} that for infinite
systems, both types of averaging  give equal results.

In the long-wavelength limit, the averages in Eq.~(\ref{D=eeffE})
are formed by those domains of coordinates in the iterative series
for $\textbf{D}(\textbf{r})$ and $\textbf{E}(\textbf{r})$ where
the inner electromagnetic field propagators reveal a singular
behavior. In other words, it is the effects of multiple
reemissions and many-particle correlations within compact groups
that contribute to $\varepsilon_{\rm eff}$. Their contributions,
in the form of statistical moments $\langle \left( \delta
\varepsilon (\textbf{r}) \right)^n \rangle$, can be singled out
from all terms of the iterative series without an in-depth
modelling of these processes. As a result,
\begin{equation}\label{sredE}
\langle\textbf{E} (\textbf{r})\rangle=\left[1+ \langle Q
(\textbf{r})\rangle \right] \textbf{E}_0,
\end{equation}
\begin{equation}\label{sredD}
\langle\textbf{D} (\textbf{r})\rangle=\varepsilon_{\rm f}\left[1-2
\langle  Q (\textbf{r})\rangle \right] \textbf{E}_0,
\end{equation}
where
\begin{equation}\label{matrix2}
Q(\textbf{r})\equiv\sum_{n=1}^{\infty}\left(-\frac{1}{3\varepsilon_{\rm
f}} \right)^n \left( \delta \varepsilon (\textbf{r}) \right)^n
\end{equation}
and ${\bf E}_0$ is the amplitude of the probing field in ${\cal
M}$.

To illustrate the general formalism, consider a dispersion of
uniform hard spheres. In this case,
\begin{equation}\label{deltaepsilon}
\delta \varepsilon (\textbf{r})=\left(\varepsilon_0
-\varepsilon_{\rm f} \right)\left(1-  \sum_{i=1}^N \Pi_i
(\textbf{r}, \Omega_i)\right)   + \left(\varepsilon_1
-\varepsilon_{\rm f} \right) \sum_{i=1}^N \Pi_i (\textbf{r},
\Omega_i),
\end{equation}
where  $\Pi_i (\textbf{r}, \Omega_i)$ is the characteristic
function for region $\Omega_i$ occupied by the $i$th sphere in a
compact group of $N\gg 1$ spheres:
\begin{equation}\label{characteristicfunction}
\Pi_i (\textbf{r}, \Omega_i)=\begin{cases}
1, & \text{if $\textbf{r} \in \Omega_i$,} \\
0, & \text{if $\textbf{r} \notin \Omega_i$.}
\end{cases}
\end{equation}
Then, by direct integration over $V$, we find from
Eqs.~(\ref{localepsilon})--(\ref{matrix2})
\cite{Sushko2009AnisPart}:
\begin{equation}\label{zagForm}
(1-c)\frac{\varepsilon_0-\varepsilon_{\rm f}}{2\varepsilon_{\rm
f}+\varepsilon_0}+ c\frac{\varepsilon_1-\varepsilon_{\rm
f}}{2\varepsilon_{\rm f}+\varepsilon_1}=\frac{\varepsilon_{\rm
eff}-\varepsilon_{\rm f}}{2\varepsilon_{\rm f}+\varepsilon_{\rm
eff}},
\end{equation}
where $c=Nv/V$ is the net volume concentration of the spheres with
individual volume $v=4\pi R^3/3$. Taking $\varepsilon_{\rm f}
=\varepsilon_0$ and $\varepsilon_{\rm f} = \varepsilon_{\rm eff}$,
we obtain, respectively,  the Maxwell-Garnett mixing rule
\cite{Maxwell1873,Maxwell1904}
\begin{equation}\label{MG}
\varepsilon_{\rm
eff}=\varepsilon_0\left(1+2c\frac{\varepsilon_1-\varepsilon_0}
{2\varepsilon_0+\varepsilon_1}\right)\bigg/\left({1-c\frac{\varepsilon_1-\varepsilon_0}{2\varepsilon_0+\varepsilon_1}}\right)
\end{equation}
and the Bruggeman mixing rule \cite{Bruggeman1935,Landauer1952}
\begin{equation}\label{Br}
(1-c)\frac{\varepsilon_0-\varepsilon_{\rm eff}}{2\varepsilon_{\rm
eff}+\varepsilon_0}+ c\frac{\varepsilon_1-\varepsilon_{\rm
eff}}{2\varepsilon_{\rm eff}+\varepsilon_1}=0.
\end{equation}

In the general case, $\varepsilon_{\rm f}$ should be treated as
unknown. We determine it consistently \cite{Sushko2016,Sushko2017}
by combining the compact-group approach with the Hashin-Shtrikman
variational principle \cite{Hashin1962}. The relevant functional
is written in terms of the local field and permittivity
distributions in ${\cal S}$. Its stationary value is equal to the
electric energy stored in ${\cal S}$ and, by definition, that in
${\cal D}$. The requirement that two different ways of
homogenization -- through the linear relation between the
induction and the field \cite{Landau1982} and through the equality
of the electric energies stored in the heterogeneous and
homogenized systems -- give the same result, enables us to
conclude that ${\cal M}$ is an imagined uniform medium with
$\varepsilon_{\rm f}=\varepsilon_{\rm eff}$ (except maybe for
metamaterials \cite{Veselago1967, Mackay2004}, not treated here).
In this case, Eqs.~(\ref{D=eeffE}), (\ref{sredE}), and
(\ref{sredD}) reduce to
\begin{equation}\label{matrix11}
\langle Q( {\bf r})\rangle=\left\langle \frac{\varepsilon ({\bf
r})-\varepsilon_{\rm {eff}}}{2\varepsilon_{\rm {eff}}+\varepsilon
({\bf r})} \right\rangle=0 .
\end{equation}

The result $\varepsilon_{\rm f}=\varepsilon_{\rm eff}$ corresponds
to the Bruggeman-type homogenization, but is not equivalent (see
\cite{Sushko2017}) to the classical Bruggeman mean-field
approximation \cite{Bruggeman1935}. It uses no extra model
considerations about the geometries and concentrations of the
constituents, permittivity distributions (except for their
piecewise continuity), and processes in the system.
Equation~(\ref{matrix11}) is actually the condition postulated for
the stochastic field  in  the SPFT
\cite{Tsang2001,Ryzhov1965,Ryzhov1970,Tamoikin1971,Tsang1981,Zhuck1994,Michel1995,Mackay2000,Mackay2001}
in order to improve the convergence of the iterative series when
solving the integral equation for $\bf E({\bf r})$.

For complete proofs of the above statements, the Reader is
referred to
\cite{Sushko2007,Sushko2009CompGroups,Sushko2009AnisPart,Sushko2017}.

\section{Systems of penetrable uniform spheres}\label{c:formalismSpheres}

\subsection{General formalism }\label{c:formalismSpheresTheory}

Suppose that the spheres in ${\cal D}$  are allowed to overlap,
the permittivity of the regions of overlap remaining equal to
$\varepsilon_1$ (see Fig.~\ref{fig:SoftBalls}). Then the
permittivity distribution in ${\cal S}$ can be modeled in
form~(\ref{localepsilon}) with $\varepsilon_{\rm f}
=\varepsilon_{\rm eff}$ and
\begin{equation}\label{profil Epsilon} \delta
\varepsilon(\textbf{r})=\Delta \varepsilon_1 \left[
1-\prod_{a=1}^N \left(1-\chi_a^{(1)}(\textbf{r}) \right) \right]+
\Delta \varepsilon_0\prod_{a=1}^N \left(
1-\chi_a^{(1)}(\textbf{r}) \right),
\end{equation}
where $\Delta \varepsilon_1 = \varepsilon_1 - \varepsilon_{\rm
eff} $, $\Delta \varepsilon_0 = \varepsilon_0 -\varepsilon_{\rm
eff}$, and $\chi_a^{(1)}(\textbf{r})=\theta(R-|\textbf{r} -
\textbf{r}_a|)$ is the characteristic function of the sphere
centered at $\textbf{r}_a$ ($\theta (x)$ is the Heaviside
function). Due to the orthogonality of the characteristic
functions at $\Delta \varepsilon_1$ and $\Delta \varepsilon_0$ in
Eq.~(\ref{profil Epsilon}), we obtain:
\begin{equation}\label{useredn profil Epsilon STEPEN}
\left \langle \left(\delta \varepsilon(\textbf{r}) \right)^s
\right \rangle=\left(\Delta \varepsilon_1 \right)^s \left[ 1-\left
\langle\prod_{a=1}^N \left( 1-\chi_a^{(1)}(\textbf{r})
\right)\right \rangle \right]  + \left   (  \Delta \varepsilon_0
\right)^s \left \langle \prod_{a=1}^N \left(
1-\chi_a^{(1)}(\textbf{r}) \right) \right \rangle.
\end{equation}

The average of the product in Eq.~(\ref{useredn profil Epsilon
STEPEN}) is the volume concentration of the host, and
\begin{multline}\label{seredniDobutky} \phi(c,\kappa) =1 - \left
\langle\prod_{a=1}^N \left(1-\chi_a^{(1)}(\textbf{r})\right)\right
\rangle =\left \langle\sum_{1\leq a \leq
N}\chi_a^{(1)}(\textbf{r})\right \rangle\\ - \left \langle\sum_{
1\leq a<b\leq N} \chi_a^{(1)}
(\textbf{r})\chi_b^{(1)}(\textbf{r})\right \rangle +\left
\langle\sum_{1\leq a<b<c \leq N} \chi_a^{(1)}(\textbf{r})
\chi_b^{(1)}(\textbf{r}) \chi_c^{(1)}(\textbf{r})\right \rangle
-\dots
\end{multline}
is the effective volume concentration of spheres
\cite{Torquato2002}. The latter is a function of the dimensionless
density $c= Nv/V$ (the ratio of the total volume of spheres to
$V$) and the hardness parameter $\kappa$, such that $\kappa =1$
corresponds to hard (mutually impenetrable) spheres and $\kappa =
0$ to ``fully'' penetrable (statistically independent) spheres.

The averages in Eq.~(\ref{seredniDobutky})  are calculated using
the $s$-sphere distribution functions $F_s(\textbf{r}_1,
\textbf{r}_2,\dots, \textbf{r}_s)$ for a given dispersion. Once
$\phi(c,\kappa)$ is known, Eq.~(\ref{matrix11}), treated as an
asymptotic series, yields the equation for $\varepsilon_{\rm
eff}$:
\begin{equation}\label{NashRezultat1}
(1-\phi(c,\kappa)) \frac{\varepsilon_0-\varepsilon_{\rm
eff}}{2\varepsilon_{\rm eff}+\varepsilon_0}+
\phi(c,\kappa)\frac{\varepsilon_1-\varepsilon_{\rm
eff}}{2\varepsilon_{\rm eff}+\varepsilon_1}=0.
\end{equation}

Rigorous calculations of $\phi(c,\kappa)$ are possible in the
limiting cases of macroscopically homogeneous and isotropic
systems of hard or fully penetrable spheres.

In the former case, $F_1({\bf r}_1) = 1$, whereas every
$F_s(\textbf{r}_1, \textbf{r}_2,\dots, \textbf{r}_s)$, $s\geq  2$,
vanishes for any sphere configuration in which at least one of the
center-to-center  distances $|\,\textbf{r}_i - \textbf{r}_j |$ is
shorter than $2R$; and so do the products of the characteristic
functions for the configurations with all $|\,\textbf{r}_i -
\textbf{r}_j |> 2R$. Then the only nonzero term after the last
equality sign in Eq.~(\ref{seredniDobutky}) is
$$ \left
\langle\sum_{1\leq a \leq N}\chi_a^{(1)}(\textbf{r})\right \rangle
=c, $$ $\phi(c,1) =c$, and Eq.~(\ref{NashRezultat1}) reduces to
the Bruggeman mixing rule~(\ref{Br}).

In the latter, the positions of spheres are statistically
independent. Then $F_s= \prod_{i=1}^s F_1(\textbf{r}_i)=1$ and
($N\gg 1$)
\begin{equation}\label{seredniDobutky2}
\left \langle \prod_{a=1}^N \left( 1-\chi_a^{(1)}(\textbf{r})
\right) \right \rangle =\left(1- \frac{v}{V} \right)^N \rightarrow
1-f,
\end{equation}
where $f=1-e^{-c}$. So, $\phi(c,0) =f$ and
Eq.~(\ref{NashRezultat1}) takes the form
\begin{equation}\label{NashRezultat}
(1-f) \frac{\varepsilon_0-\varepsilon_{\rm eff}}{2\varepsilon_{\rm
eff}+\varepsilon_0}+ f\frac{\varepsilon_1-\varepsilon_{\rm
eff}}{2\varepsilon_{\rm eff}+\varepsilon_1}=0.
\end{equation}

In deriving Eq.~(\ref{NashRezultat}), the fact of sphericity of
particles (the explicit form of $\chi_a^{(1)}(\textbf{r})$) was
not used. It follows immediately that Eq.~(\ref{NashRezultat})
should also hold for macroscopically homogeneous and isotropic
systems prepared by embedding $N\gg 1$ freely overlapping regions
of arbitrary shape, with individual permittivities $\varepsilon_1$
and volumes $v$, into a host of permittivity $\varepsilon_0$.
Moreover, Eq.~(\ref{NashRezultat}) should remain valid for
polydisperse mixtures of such regions, having differing shapes and
volumes, $v_\alpha$, provided the number $N_\alpha$ of regions of
each sort $\alpha = 1, 2,\dots, q$ is statistically large:
$N_\alpha\gg 1$. The effective volume concentration of such
regions $f=1-\exp\left(-\Sigma_{\alpha =1}^{q} N_\alpha v_\alpha/V
\right)$ and the volume concentration of the host is $1-f$.

\subsection{Systems of soft particles}\label{c:formalismSpheresSoft}
For dispersions of soft spheres, with $0<\kappa<1 $, the
determination of $\kappa$ and theoretical calculation of
$\phi(c,\kappa)$ are nontrivial problems. One way is to define
$\kappa$ through the pair distribution function
$F_2(\textbf{r}_1,\textbf{r}_2)=F_2(|\textbf{r}_1 -
\textbf{r}_2|)$ by the relation \cite{Blum1979,Salacuse1982}
$$F_2(r) =1-\kappa, \quad r <2R.
$$
Assuming the direct correlation function and the pair potential to
be zero for $r >2R$, one can use the Percus-Yevick approximation
for $F_2(r)$ and the generalized Kirkwood superposition
approximation for $F_s(\textbf{r}_1, \textbf{r}_2,\dots,
\textbf{r}_s)$ at $s > 2$ to obtain the estimate
\cite{Rikvold1985}
\begin{equation}\phi(c,\kappa) \approx
\sum^{\infty}_{n=1}\frac{(-1)^{n+1}c^n}{n!}(1-\kappa)^{{n(n-1)}/{2}}.
\label{softconc}
\end{equation}
This, in particular, reproduces the above limiting values:
$\phi(c,0) =f$, $\phi(c,1) =c$. The equation for $\varepsilon_{\rm
eff}$ still has form (\ref{NashRezultat1}).

Note that the concept of hard particles implies the existence of a
sharp change in density at the particle-host interface. In fact,
not many interfaces fulfill this condition. Particles may contain
pores, be gel-type penetrable or somewhat deformable, have
``hairy'' adjacent polymer layers, etc. Given permittivity data
for a system of such ``soft'' particles, we can use
Eq.~(\ref{NashRezultat1}) to recover $\phi(c,\kappa)$  and then
Eq.~(\ref{softconc}) to estimate  $\kappa$. In this way, the
effective hardness of particles can be defined operationally.

\subsection{Dilute dispersions }\label{c:formalismSpheresComparisonDiluted}
It is of crucial importance to verify  the applicability of our
theory to diluted dispersions, for which  a number of reliable
results exist in the literature and, on the other hand, the
concept of compact groups may seem most vulnerable.

In the limit $\phi(c,\kappa)\to 0$, Eq.~(\ref{NashRezultat1}) can
be represented as
\begin{equation}\label{Asymptotic}
\frac{\varepsilon_{\rm eff}}{\varepsilon_0}=1+3\beta
\phi(c,\kappa) +\left(3\beta^2+6\beta^3
\right)\phi^2(c,\kappa)+O\left(\beta^3\phi^3(c,\kappa)\right),
\end{equation}
where $\beta =\left(\varepsilon_1-\varepsilon_0
\right)/\left(\varepsilon_1+2\varepsilon_0 \right)$. For
$\phi(c,\kappa)$ given by Eq.~(\ref{softconc}) and $c\to 0$,
Eq.~(\ref{Asymptotic}) takes the form
\begin{equation}\label{AsymptoticPenetr}
\frac{\varepsilon_{\rm eff}}{\varepsilon_0}=1+ 3 \beta c +
 \left[ -\frac{3}{2} (1-\kappa)\beta + 3\beta^2 + 6\beta^3\right] c^2 +
O\left(\beta c^3\right).
\end{equation}

The first two addends on the right of Eq.~(\ref{AsymptoticPenetr})
give the classical Maxwell-Garnett result
\cite{Maxwell1873,Maxwell1904}. The next addend is a
$c^2$-correction. For a dilute hard-sphere gas ($\kappa=1$), its
leading term $3\beta^2c^2$ agrees with those in Jeffrey's result
\cite{Jeffrey1973}
\begin{equation}\label{AsymptoticJeffrey}
\frac{\varepsilon_{\rm eff}}{\varepsilon_0}=1+ 3 \beta c +
 \left[3\beta^2 + 6A(\beta)\beta^3\right] c^2 +
O\left(c^3\right)
\end{equation}
and result \cite{Felderhof1982a}; the other term, $\sim \beta^3
c^2$, differs from its counterparts in
\cite{Jeffrey1973,Felderhof1982a} by a $\beta$-dependent
coefficient $A(\beta)<0.252$. For low-contrast dispersions with
$\delta=|\varepsilon_1 -\varepsilon_0|/\varepsilon_0\ll 1$, all
the results agree, to $O(\delta^2)$, with one another and with
relation \cite{Landau1982}
$$\varepsilon_{\rm eff} = \overline{\varepsilon} - \frac{1}{3\overline{\varepsilon}}
\overline{(\varepsilon - \overline{\varepsilon})^2},$$ valid to
$O(\delta^2)$ for any mixture in which the local variations of
permittivity are weak.

If the spheres are penetrable ($0\leq \kappa <1$), the volume
concentration $\phi(c,\kappa)$ is a  more preferable parameter
than $c$. Through order $c^2$, $\phi(c,\kappa)
=c-\frac{1}{2}(1-\kappa)c^2\equiv \phi_2$ and
Eq.~(\ref{Asymptotic}) approaches Torquato's result
\cite{Torquato1984,Torquato1985,Torquato1985b}
\begin{equation}\label{AsymptoticTorq}
\frac{\varepsilon_{\rm eff}}{\varepsilon_0}=1+3\beta \phi_2
+\left[3\beta^2+6(0.21068+0.35078(1-\kappa)) \beta^3
\right]\phi_2^2,
\end{equation}
except for the numerical factor in front of $\beta^3 \phi_2^2$.
Note that for dispersions of fully penetrable spheres,
Eq.~(\ref{AsymptoticTorq}) is expected to be a good approximation
provided $\phi_2^2 < 0.2$ \cite{Torquato1985}.

So, the agreement of Eq.~(\ref{Asymptotic}) with results
\cite{Felderhof1982a,Torquato1984,Torquato1985,Jeffrey1973} for
dilute dispersions with low $\delta$ is exact through order
$\beta^2 \phi_2^2$. At the same time, the $O(\beta^3
\phi_2^2)$-corrections are of greater magnitude in our theory.
This fact can be interpreted as a manifestation of  many-particle
polarization and correlation effects, coming into play as $\delta$
increases and effectively taken into account by
Eq.~(\ref{Asymptotic}).

\subsection{Comparison with numerical experiment}\label{c:formalismSpheresComparison}

To further test our theory, we contrast it with differential
mixing equation \cite{Jylha2007}, proposed for mixtures with
wide-ranging permittivity contrast $k =
\varepsilon_1/\varepsilon_0$, and numerical $\nu$-model
\cite{Karkainen2001}
\begin{equation}\label{nuModel}
\frac{\varepsilon_{\rm eff}-\varepsilon_0}{\varepsilon_{\rm
eff}+2\varepsilon_0+\nu(\varepsilon_{\rm eff}-\varepsilon_0)}
=f\frac{\varepsilon_1-\varepsilon_0}{\varepsilon_1+2\varepsilon_0+\nu(\varepsilon_{\rm
eff}-\varepsilon_0)},
\end{equation}
derived as a  fit to the set of over 4000 simulation results for
$k$ raging from $1/102$ to $102$. The dimensionless fitting
function $\nu = \nu (k,f)$ was proposed to be
\begin{equation}\label{nu} \nu (f,k) =
\begin{cases}
(1.27+1.43\,e^{-0,048k})f^2+(-2.76-0.9\,e^{-0.043k})f+2.35,&k>1,\\
1.06f^2+(-1.23+0.44\,e^{-5.95k})f+1.7,&k<1.
\end{cases}
\end{equation}
The permittivity $\varepsilon_{\rm eff}$  was determined  by
calculating the electrostatic field energy of samples placed in a
homogeneous electric field. The computational domain, of size
 $100\times 100\times 100=1000000$ cells, was restricted
with periodic boundary conditions. The  potentials were different
on the faces perpendicular to the field and the same on the other
faces.  The diameter of  randomly positioned spheres was 20 cells.

\begin{figure}[!tbp]
  \centering
  \begin{minipage}[t]{0.45\textwidth}
    \includegraphics[width=\textwidth]{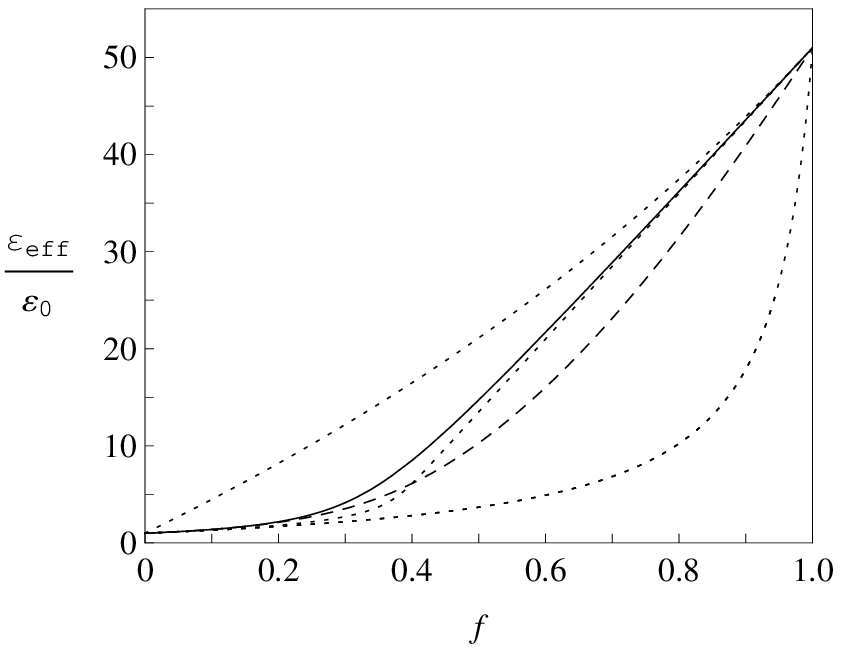}
    \caption{\label{fig:Comparison1Dif}{\normalsize $\varepsilon_{\rm eff}/\varepsilon_0$ versus $f$ for a
dispersion of fully penetrable uniform spheres according to
Eq.~(\ref{NashRezultat}) (solid curve),  $\nu$-model
(\ref{nuModel}) (dashed one), and differential mixing equation
\cite{Jylha2007} at $k=51$ (middle dotted one); the outermost
dotted curves represent the Hashin-Shtrikman bounds
\cite{Hashin1962}.}}
  \end{minipage}
  \hfill
  \begin{minipage}[t]{0.45\textwidth}
    \includegraphics[width=\textwidth]{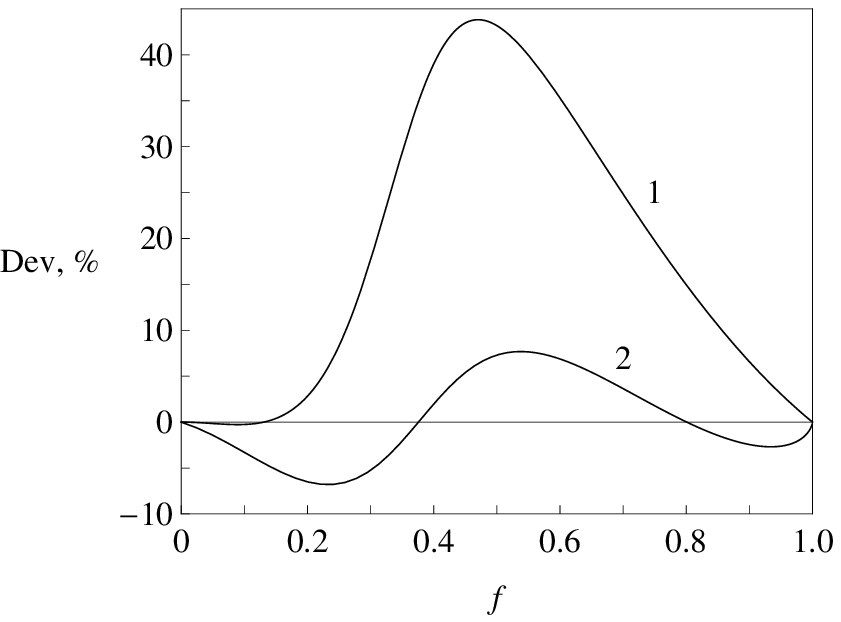}
    \caption{\label{fig:Deviation12}{\normalsize The relative deviations of $\varepsilon_{\rm eff}$
    given by  Eq.~(\ref{NashRezultat}) and that by Eq.~(\ref{equatDor3}) at $a=0.93$, $\varepsilon_2/\varepsilon_0=5$
    (curves 1 and 2, respectively) from $\varepsilon_{\rm eff}$ given by $\nu$-model (\ref{nuModel}) as functions of
$f$; $ k=51$. }}
  \end{minipage}
\end{figure}

The finite difference approximation was used in
\cite{Karkainen2001} to solve the equation for potential. The
local permittivity value in the difference equation was ambiguous
when the cubic grid cell included the surface of a sphere. In such
cases, it was taken to be equal to certain weighted averages of
$\varepsilon_0$ and $\varepsilon_1$. That is, a third phase, with
poorly defined permittivity $\varepsilon_2$, was actually
introduced.

The results for $\varepsilon_{\rm eff }$  obtained in the
framework of our approach, equation \cite{Jylha2007}, and
$\nu$-model (\ref{nuModel}) at $k=51$ are shown in
Fig.~\ref{fig:Comparison1Dif}; all are within the Hashin-Shtrikman
bounds. The relative deviation ${\rm Dev} = \left(
\varepsilon_{\rm eff}^{\rm our} -\varepsilon_{\rm
eff}^{\nu}\right)/\varepsilon_{\rm eff}^{\nu}\cdot 100\%$ between
the results is shown in Fig.~\ref{fig:Deviation12}. For $f \gtrsim
0.2$, our theory predicts higher values for $\varepsilon_{\rm
eff}$; a possible reason is that the uncontrolled $\varepsilon_2$
took intermediate values between $\varepsilon_0$ and
$\varepsilon_1$. If so, the appearance of the maximum in
Fig.~\ref{fig:Deviation12} is explained readily: as $f$ increases,
the net volume and, therefore, the accumulating effect of the
interfacial layers should first increase and then, when the
spheres begin to overlap considerably, decrease. That fact that
our theory is in rather good agreement with analytical
differential mixing equation \cite{Jylha2007} (see
Fig.~\ref{fig:Comparison1Dif}) seems to support our suggestion.

Note that a statistically small number of spheres and anisotropy
of the computing domain may also be factors contributing to ${\rm
Dev.}$

\section{Systems  of  penetrable heterogeneous  spheres}\label{c:inhomogeneusParticles}
\subsection{Equation for $\varepsilon_{\rm eff}$}\label{c:inhomogeneusParticlesEquation}

The preceding analysis naturally leads to the problem of finding
$\varepsilon_{\rm eff}$ for systems  of heterogeneous particles.
For hard particles, its formal solution within the compact-group
approach was obtained in
\cite{Sushko2009CompGroups,Sushko2009AnisPart,Sushko2017}. Here,
we analyze the case where: 1) the particles are fully penetrable
isotropic spheres embedded into an isotropic and homogeneous host
of permittivity $\varepsilon_0$; 2) their permittivity profile is
described by a piecewise-continuous radial function $e=e(r)$; 3)
the local permittivity value $\varepsilon({\bf r})$ in the system
is determined, according to \cite{Sushko2016,Sushko2018}, by the
distance $l\equiv\min_{a} |\,{\bf r}-{\bf r}_a |$ from the point
of interest ${\bf r}$ to the center of the nearest sphere. An
example of such a system is shown in
Fig.~\ref{fig:SoftLayeredBalls}.

Following the line of reasoning \cite{Sushko2016,Sushko2018},
suppose first that every sphere, of radius $R$, consists of $M$
concentric spherical layers with outer radii $R_j$ and constant
permittivities $\varepsilon_j$, $j= 1, 2, \dots, M$
($R_1<R_2<\dots <R_M\equiv R$). According to the above
assumptions, the local permittivity is given by
\begin{equation} \varepsilon({\bf r})=\begin{cases}
\varepsilon_1, & {\text{if} }
\quad \,\,l<R_1,\\
\varepsilon_j, & {\text{if} }\quad  R_{j-1}<l<R_j, (2\leq j \leq M),\\
\varepsilon_0, & {\text{if} }\quad \,\, l>R_M.
\end{cases} \label{distr1}
\end{equation}
Introducing the characteristic functions $\chi_a^{(i)}({\bf r})
=\theta\left(R_i -|\,{\bf r}-{\bf r}_a|\right)$ for spheres with
centers at points ${\bf r}_a$ and radii $R_i$, we rewrite
Eq.~(\ref{distr1}) in form (\ref{localepsilon}) with
$\varepsilon_{\rm f}=\varepsilon_{\rm eff}$ and
\begin{multline}\label{perm00Dor}
\delta\varepsilon({\bf r})=
\Delta\varepsilon_1\left[1-\prod\limits_{a=1}^N
\left(1-\chi_a^{(1)}\left({\bf r}\right)\right)\right]\\+
\sum\limits_{j=2}^M
\Delta\varepsilon_j\left[1-\prod\limits_{a=1}^N
\left(1-\chi_a^{(j)}\left({\bf r}\right)\right)\right]
\prod\limits_{b=1}^N \left(1-\chi_b^{(j-1)}\left({\bf
r}\right)\right)+ \Delta\varepsilon_0 \prod\limits_{b=1}^N
\left(1-\chi_b^{(M)}\left({\bf r}\right)\right),
\end{multline}
where $\Delta\varepsilon_0=\varepsilon_0-\varepsilon_{\rm eff}$
and $\Delta\varepsilon_j=\varepsilon_j-\varepsilon_{\rm eff}$. By
doing so, we change from a given system, ${\cal D^*}$, to the
associated system, ${\cal S^*}$, used for the homogenization of
${\cal D^*}$ (as discussed in sections~\ref{c:introduction} and
\ref{c:compactgroups}).

The moments of $\delta\varepsilon({\bf r})$ are
\begin{multline}\label{perm22Dor}
\langle\left(\delta\varepsilon({\bf r})\right)^s\rangle =
f(n,R_1)\left(\Delta\varepsilon_1\right)^s
+\sum\limits_{j=2}^M\left[f(n,R_j)-f(n,R_{j-1})\right]\left(\Delta\varepsilon_j\right)^s
+\left[1-f(n,R_M)\right]\left(\Delta\varepsilon_0\right)^s,
\end{multline}
where
\begin{equation}\label{perm33Dor}
f(n,R_j) =1-\left\langle\prod\limits_{a=1}^N \left(1-\chi_a^{(j)}
\left({\bf r}\right)\right) \right\rangle=1-e^{-4 \pi nR^3_j/3}
\end{equation}
and $n=N/V$ is the particle number density.

Now, passing to the limits $M \to \infty$, $| R_j-R_{j-1} |\to~0$
and taking into account the differentiability of $f(n,r)$ in $r$,
we can generalize Eq.~(\ref{perm22Dor}) to the case where the
sphere's permittivity profile is a piecewise-continuous function
$e=e(r)$:
\begin{equation}\label{perm222Dor}
\langle\left(\delta\varepsilon({\bf r})\right)^s\rangle=
\int\limits_0^R\frac{\partial f(n,r)}{\partial r}\left[\Delta
e(r)\right]^s dr
+\left(1-f(n,R)\right)\left(\Delta\varepsilon_0\right)^s,
\end{equation}
where $\Delta e(r) = e(r)-\varepsilon_{\rm eff}$. Then
Eqs.~(\ref{perm33Dor}) and (\ref{matrix11}) give the desired
equation for $\varepsilon_{\rm eff}$:
\begin{equation} \label{equatDor1}
e^{-4 \pi nR^3/3}\frac{\varepsilon_0 -\varepsilon_{\rm
eff}}{2\varepsilon_{\rm eff}+\varepsilon_0} +4 \pi
n\int\limits_0^R
 r^2 e^{-4 \pi nr^3/3}\frac{e (r)
-\varepsilon_{\rm eff}}{2\varepsilon_{\rm eff}+e(r)}\,dr =0.
\end{equation}
Changing to the dimensionless variable $u=r/R$, we can represent
it as
\begin{equation} \label{equatDor1a}
e^{-c}\frac{\varepsilon_0 -\varepsilon_{\rm
eff}}{2\varepsilon_{\rm eff}+\varepsilon_0} + 3 c \int\limits_0^1
 u^2 e^{-cu^3}\frac{e (u)
-\varepsilon_{\rm eff}}{2\varepsilon_{\rm eff}+e(u)}\,du =0.
\end{equation}

The analogous equation for a dispersion of hard spheres with a
piecewise-continuous permittivity profile $e=e(r)$ reads
\cite{Sushko2009CompGroups,Sushko2009AnisPart,Sushko2017}
\begin{equation} \label{equatDor1b}
(1-c)\frac{\varepsilon_0 -\varepsilon_{\rm eff}}{2\varepsilon_{\rm
eff}+\varepsilon_0} + 3 c \int\limits_0^1
 u^2 \frac{e (u)
-\varepsilon_{\rm eff}}{2\varepsilon_{\rm eff}+e(u)}\,du =0.
\end{equation}

\subsection{Comparison with analytical results for inhomogeneous spheres}

For low values of $k$ and $c$, both electromagnetic interactions
and spatial correlations are small. In this case, the
Maxwell-Garnett- and Bruggeman-type approaches should give close
results. This fact was used in \cite{Sushko2017} to test
Eq.~(\ref{equatDor1b}) by contrasting its solutions with
analytical results \cite{Sihvola1989} for mixtures of hard spheres
with continuous radial permittivity profiles. The quasistatic
polarizability of spheres and then $\varepsilon_{\rm eff}$ were
calculated in \cite{Sihvola1989} using (a) the internal field
method (one finds the dipole moment by integrating the product of
the field and the permittivity over the sphere's volume) and (b)
the external field method (one finds the field perturbation due to
the sphere and then the amplitude of an equivalent dipole). Both
methods gave the same results, free of ambiguities typical of the
above computer simulations.

\begin{figure}[!tbp]
  \centering
  %\begin{minipage}[t]{0.45\textwidth}
    \includegraphics[width=75mm]{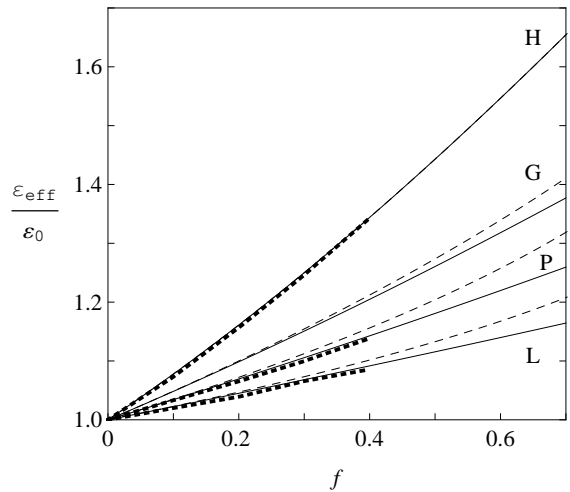}
    \caption{\label{fig:ComparisonSihvola} {\normalsize $\varepsilon_{\rm eff}/\varepsilon_0$ versus
$f$ for dispersions of continuously inhomogeneous hard (solid
lines, $f=c$, \cite{Sushko2017}) and fully penetrable (dashed
lines, $f=1-e^{-c}$) dielectric spheres according to
Eqs.~(\ref{equatDor1b}) and (\ref{equatDor1a}), respectively; H,
L, P and G specify different types of the permittivity profiles
(see text). The dotted lines are analytical results
\cite{Sihvola1989} for inhomogeneous hard spheres with the same
profiles H, L and P.}}
  %\end{minipage}
\end{figure}

Figure~\ref{fig:ComparisonSihvola} represents  our results
obtained by Eqs.~(\ref{equatDor1b}) and (\ref{equatDor1a}) for
$\varepsilon_{\rm eff}$ of dispersions of inhomogeneous hard and
fully penetrable dielectric particles embedded in a medium of
permittivity $\varepsilon_0=1$ and having the following
permittivity profiles ($0 \leq u \leq 1$): homogeneous $e(u)= 2\,
\varepsilon_0$ (denoted as H); linear $e(u)= \varepsilon_0 (2-u)$
(L); parabolic $e(u)= \varepsilon_0 \left(2-u^2\right)$ (P);
Gaussian $e(u) = \varepsilon_0 \left(1+e^{-u^2}\right)$ (G; this
is different from the Gaussian packet used in \cite{Sihvola1989}
since we consider spheres with sharp boundaries). The agreement of
our results \cite{Sushko2017} with analytical results
\cite{Sihvola1989} for inhomogeneous hard spheres is surprisingly
good. It is also seen that the penetrability of spheres with the
same profiles causes $\varepsilon_{\rm eff}$ to increase, as
compared to $\varepsilon_{\rm eff}$ for the systems of hard
spheres. This increase is caused by the penetrable spheres'
internal regions which have higher  permittivities and gradually
come into play as $f$ becomes greater. However, because of the low
permittivity contrasts between the regions, its relative magnitude
is not very high ($ < 3.8, \,4.8$ and $2.4 \%$ for L, P and G,
respectively; $f \in[0,0.7]$).  No increase is observed for
homogeneous penetrable spheres, as expected.

\subsection{Further comparison with the numerical experiment} \label{c:further}

For fully penetrable two-layer spheres, with the inner-sphere
radius $aR$ ($a\leq 1$) and permittivity $\varepsilon_1$  and the
outer-sphere radius $R$ and permittivity $\varepsilon_2$,
Eq.~(\ref{equatDor1a}) can be represented as
\begin{equation} \label{equatDor3}
(1-f)\frac{\varepsilon_0 -\varepsilon_{\rm eff}}{2\varepsilon_{\rm
eff}+\varepsilon_0} +\left[1- (1-f)^{a^3}
\right]\frac{\varepsilon_1 -\varepsilon_{\rm
eff}}{2\varepsilon_{\rm eff}+\varepsilon_1}+  \left[ (1-f)^{a^3}
-(1-f)\right] \frac{\varepsilon_2 -\varepsilon_{\rm
eff}}{2\varepsilon_{\rm eff}+\varepsilon_2}=0.
\end{equation}
Recalling our remark (see  subsection
\ref{c:formalismSpheresComparison}) that the procedure used in
\cite{Karkainen2001} to determine the local permittivity value
near the sphere-host interfaces is equivalent to the introduction
of a third phase, we expect that Eq.~(\ref{equatDor3}) is capable
of reproducing the $\nu$-model results within the accuracy of the
fitting procedure employed in \cite{Karkainen2001} to obtain
Eqs.~(\ref{nuModel}) and (\ref{nu}). Based on Figs.~6 and 7 in
\cite{Karkainen2001}, we estimate that accuracy to be no better
than $10\%$ for $f \in [0.20, 0.65]$.

%\begin{figure}[h]
%\noindent \centering
%\includegraphics[width=80mm]{Comparison2.eps}
%\caption{$\varepsilon_{\rm eff}/\varepsilon_0$ versus $f$ for a
%system of fully penetrable two-layer spheres according to
%(\ref{equatDor3}) at $a=0.93$, $\varepsilon_2/\varepsilon_0 =5$
%(solid curve); $k=51$. The dots are simulation results
%\cite{Karkainen2001} for uniform spheres with $k=51$; the dashed
%line represents  the corresponding $\nu$-model result.}
%\label{fig:Comparison2}
%\end{figure}

%\begin{figure}[h]
%\noindent \centering
%\includegraphics[width=80mm]{Deviation2.eps}
%\caption{The relative deviation of $\varepsilon_{\rm eff}$ given
%by (\ref{equatDor3}) at $a=0.93$, $\varepsilon_2/\varepsilon_0=5$
%from that given by $\nu$-model (\ref{nuModel}) as a function of
%$f$; $ k=51$. } \label{fig:Deviation2}
%\end{figure}

\begin{figure}[!tb]
  \centering
  \begin{minipage}[t]{0.45\textwidth}
    \includegraphics[width=\textwidth]{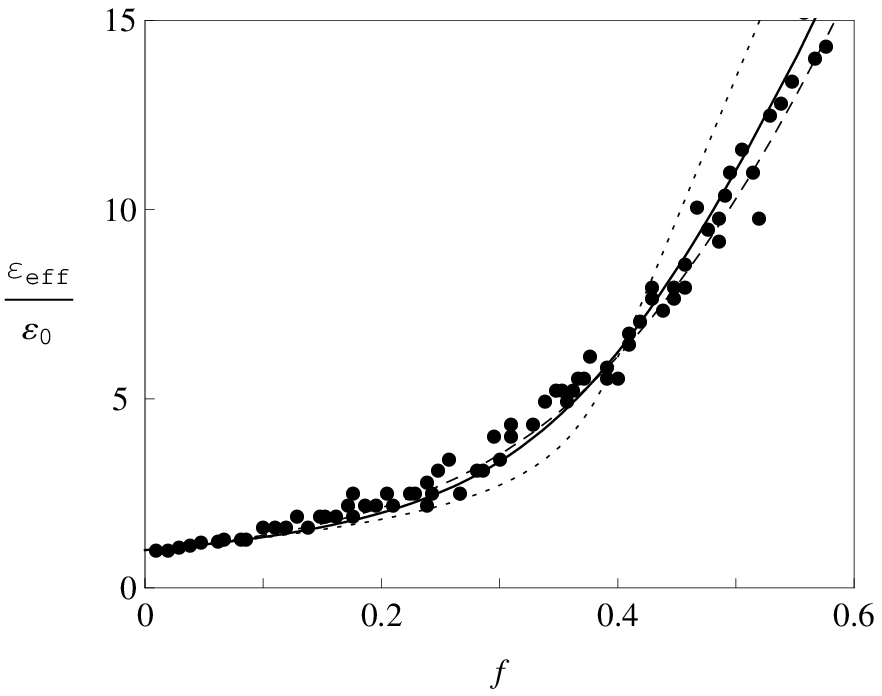}
    \caption{\label{fig:Comparison2Dif1}{\normalsize $\varepsilon_{\rm eff}/\varepsilon_0$ versus $f$ for a
dispersion of fully penetrable two-layer spheres according to
Eq.~(\ref{equatDor3}) at $a=0.93$, $\varepsilon_2/\varepsilon_0
=5$ (solid curve); $k=51$. The dots are simulation results
\cite{Karkainen2001} for uniform spheres with $k=51$; the dashed
and dotted lines correspond to $\nu$-model (\ref{nuModel}) and
differential mixing equation \cite{Jylha2007}, respectively.}}
  \end{minipage}
  \hfill
  \begin{minipage}[t]{0.45\textwidth}
    \includegraphics[width=\textwidth]{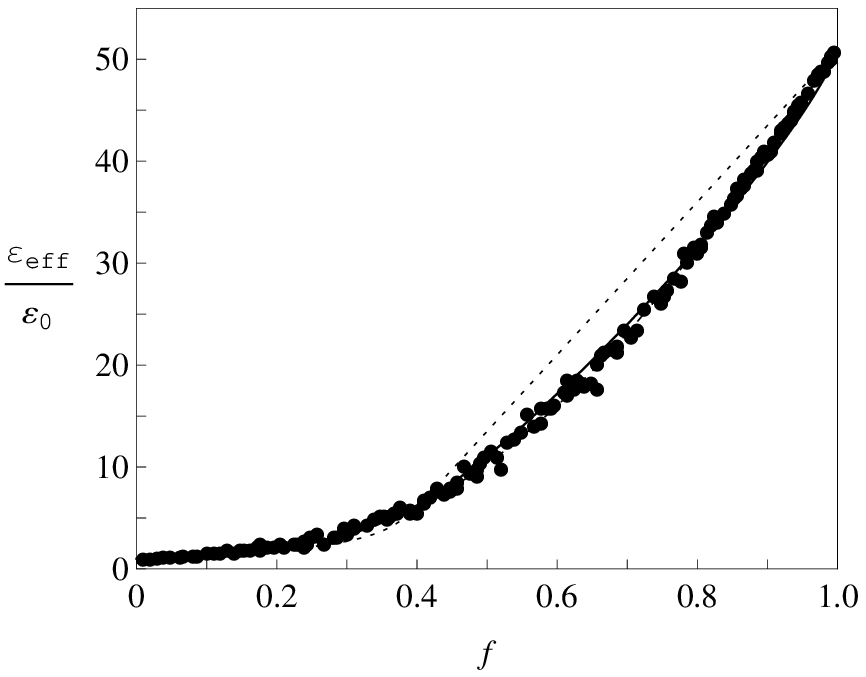}
    \caption{ \label{fig:Comparison2Dif2}{\normalsize The same as in Fig.~\ref{fig:Comparison2Dif1}, but for the
    entire range of $f$. }}
  \end{minipage}
\end{figure}

The results predicted by model (\ref{equatDor3}) for $a=0.93$,
$\varepsilon_2/\varepsilon_0=5$, and $k=51$, simulations
\cite{Karkainen2001}, $\nu$-model (\ref{nuModel}), and
differential mixing equation \cite{Jylha2007} are compared in
Figs.~\ref{fig:Comparison2Dif1}, \ref{fig:Comparison2Dif2}, and
\ref{fig:Deviation12}. The maximum deviation of our theory from
the $\nu$-model does not exceed $7.5 \%$ on the entire interval $f
\in [0,1]$. For narrower intervals of $f$, it can be greatly
reduced by minor variations in $a$ and $\varepsilon_2$.

\begin{figure}[!tbp]
  \centering
  \begin{minipage}[t]{0.45\textwidth}
    \includegraphics[width=\textwidth]{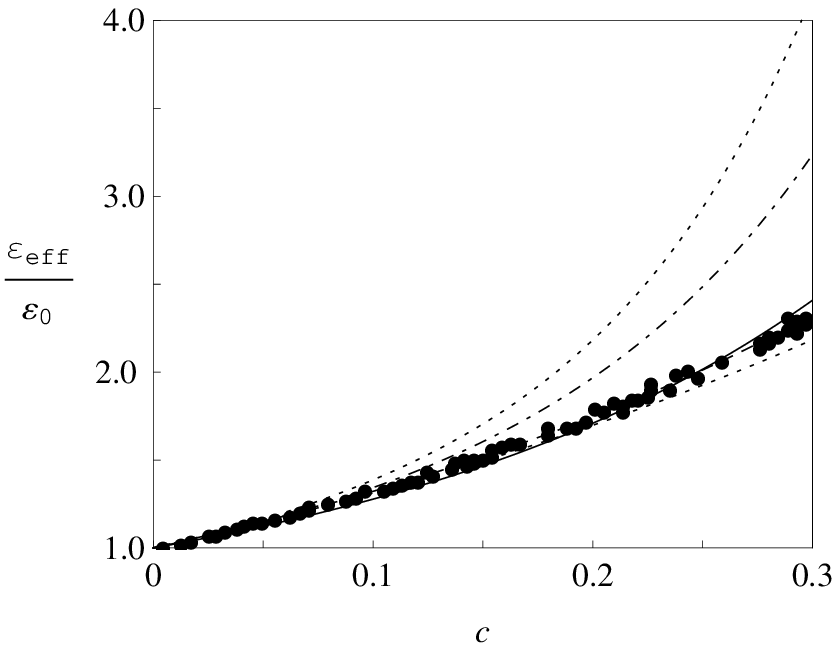}
    \caption{\label{fig:HardBallsComparison}{\normalsize $\varepsilon_{\rm eff}/\varepsilon_0$ versus $c$ for a
dispersion of hard two-layer spheres according to
Eq.~(\ref{equatDor3Hard}) at $a=0.93$,
$\varepsilon_2/\varepsilon_0=5$ (dash-dot curve), $a=0.82$,
$\varepsilon_2/\varepsilon_0 =4.1$ and $a=0.79$,
$\varepsilon_2/\varepsilon_0=5$ (solid curves, practically
indistinguishable); $k=51$. The dots are simulation results
\cite{Karkainen2001}, the dashed curve is the $\nu$-model result
at $\nu =0.30$, and the dotted curves (upper and lower,
respectively) are the Bruggeman and Maxwell-Garnett results, all
for a dispersion of hard uniform spheres with $k=51$.}}

  \end{minipage}
  \hfill
  \begin{minipage}[t]{0.45\textwidth}
    \includegraphics[width=\textwidth]{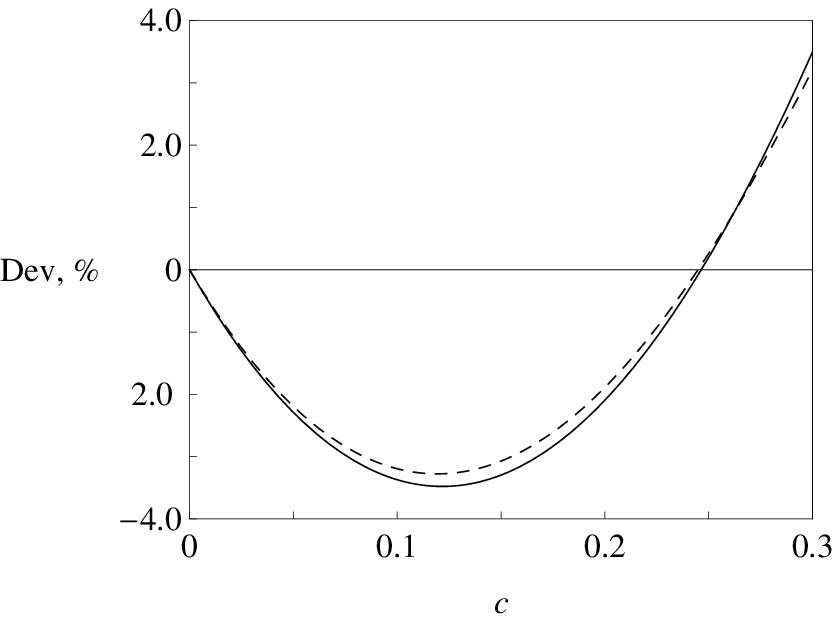}
    \caption{\label{fig:HardBallsDeviation} {\normalsize The relative deviations of $\varepsilon_{\rm eff}$ given
by Eq.~(\ref{equatDor3Hard}) at $a=0.82$,
$\varepsilon_2/\varepsilon_0=4.1$ (solid curve) and that at
$a=0.79$, $\varepsilon_2/\varepsilon_0=5$ (dashed curve) from
 the $\nu$-model
result at $\nu =0.30$  as functions of $c$; $ k=51$. }}
  \end{minipage}
\end{figure}

%\subsection{Case of hard spheres}
The authors of \cite{Karkainen2001} also studied mixtures of hard
dielectric spheres with volume fractions $c <0.3$. The results
converged rather poorly when the grid step size was reduced. To
improve their convergence, the local permittivity near the
sphere-host interfaces was found by using the minimum estimation
within the numerical technique employed, rather than the average
of limit estimations. The final results for $\varepsilon_{\rm
eff}$ at $k=51$ are presented in
Fig.~\ref{fig:HardBallsComparison} and can be described by the
$\nu$-model with $\nu \approx 0.3$, which is rather close the
Maxwell–Garnett model ($\nu = 0$). They give us another
opportunity to test our approach.

%\begin{figure}[h]
%\noindent \centering
%\includegraphics[width=80mm]{HardBallsComparison.eps}
%\caption{$\varepsilon_{\rm eff}/\varepsilon_0$ versus $c$ for a
%system of hard two-layer spheres according to
%(\ref{equatDor3Hard}) at $a=0.93$, $\varepsilon_2/\varepsilon_0=5$
%(dash-dot curve), $a=0.82$, $\varepsilon_2/\varepsilon_0 =4.1$ and
%$a=0.79$, $\varepsilon_2/\varepsilon_0=5$ (solid curves,
%practically indistinguishable); $k=51$. The dots are simulation
%results \cite{Karkainen2001}, the dashed curve is the $\nu$-model
%result at $\nu =0.30$, and the dotted curves are the Bruggeman
%(upper curve) and Maxwell-Garnett (lower curve) results, all for a
%system of hard uniform spheres with $k=51$. }
%\label{fig:HardBallsComparison}
%\end{figure}

%\begin{figure}[h]
%\noindent \centering
%\includegraphics[width=80mm]{HardBallsDeviation.eps}
%\caption{The relative deviations of $\varepsilon_{\rm eff}$ given
%by (\ref{equatDor3Hard}) at $a=0.82$,
%$\varepsilon_2/\varepsilon_0=4.1$ (solid curve) and that at
%$a=0.79$, $\varepsilon_2/\varepsilon_0=5$ (dashed curve) from
% the $\nu$-model
%result at $\nu =0.30$  as functions of $c$; $ k=51$. }
%\label{fig:HardBallsDeviation}
%\end{figure}

Consider a dispersion of hard two-layer  spheres, with the
inner-sphere radius $aR$ ($a\leq 1$) and permittivity
$\varepsilon_1$  and the outer-sphere radius $R$ and permittivity
$\varepsilon_2$. According to Eq.~(\ref{equatDor1b}), the equation
for its $\varepsilon_{\rm eff}$ is
\begin{equation} \label{equatDor3Hard} (1-c)\frac{\varepsilon_0
-\varepsilon_{\rm eff}}{2\varepsilon_{\rm eff}+\varepsilon_0}
+a^3c\,\frac{\varepsilon_1 -\varepsilon_{\rm
eff}}{2\varepsilon_{\rm eff}+\varepsilon_1}+  \left(1 -
a^3\right)c\,\frac{\varepsilon_2 -\varepsilon_{\rm
eff}}{2\varepsilon_{\rm eff}+\varepsilon_2}=0.
\end{equation}
In our view, the just-mentioned choice of the local permittivity
in  \cite{Karkainen2001} is equivalent to a decrease in
$\varepsilon_2$,  or  $a$, or both, as compared to those for the
model of penetrable two-layer  spheres. In particular,  for two
sets of parameters, $a=0.82$, $\varepsilon_2/\varepsilon_0 =4.1$
and $a=0.79$, $\varepsilon_2/\varepsilon_0=5$,
Eq.~(\ref{equatDor3Hard}) reproduces the $\nu$-model result for
hard uniform spheres with an accuracy of no worse than 3.5 \% (see
Fig.~\ref{fig:HardBallsComparison} and
\ref{fig:HardBallsDeviation}). This  fact is another evidence for
the efficiency and consistency of our approach.

\section{Conclusion}\label{c:conclusion}

The main results of this study can be summarized as follows.
\begin{enumerate}
\item  Using the compact-group approach \cite {Sushko2007,
Sushko2009CompGroups, Sushko2009AnisPart,Sushko2016,Sushko2017},
complemented by the Hashin-Shtrikman variational theorem
\cite{Hashin1962}, we developed a many-particle theory for finding
the effective quasistatic permittivity $\varepsilon_{\rm eff}$ of
dispersions of dielectric particles with different degrees of
penetrability. According to it:
\begin{enumerate}
\item  a dispersion to be homogenized is dielectrically equivalent
to a macroscopically homogeneous and isotropic system prepared by
embedding the constituents of the real dispersion into an imagined
medium having the looked-for permittivity (Bruggeman-type
homogenization); \item the governing equation for
$\varepsilon_{\rm eff}$ is obtained from Eq.~(\ref{matrix11}) by
summing up the statistical moments for the deviations of the local
permittivity values in the model system from $\varepsilon_{\rm
eff}$. These moments are determined by the properties of the
dispersion's constituents, such as their geometric parameters,
permittivity profiles, degree of penetrability, etc.; \item
$\varepsilon_{\rm eff}$ is  found from the governing equation as a
functional of the dispersion's constituents’ permittivity profiles
and volume concentrations, the latter being expressed through the
statistical averages of certain products of the particles'
characteristic functions.
\end{enumerate}
\item  The theory was applied to dispersions of spheres with a
piecewise-continuous radial permittivity profile under the
suggestion that for overlapping spheres, the local permittivity
value is determined by the shortest distance from the point of
interest to their centers. In this case, $\varepsilon_{\rm eff}$
satisfies certain integral relations, which were analyzed in
detail. \item The efficiency of the theory was demonstrated by
contrasting its results with:
\begin{enumerate}
\item rigorous calculations
\cite{Felderhof1982a,Torquato1984,Torquato1985,Jeffrey1973} for
dilute dispersions of uniform spheres with different degrees of
penetrability; \item analytical calculations \cite{Sihvola1989}
for low-contrast dispersions ($1 \leq \varepsilon_1/\varepsilon_0
\leq 2$) of hard spheres with different permittivity profiles and
volume concentrations $c \in [0,0.4]$; \item differential mixing
equation \cite{Jylha2007} for high-contrast
($\varepsilon_1/\varepsilon_0 \gg 1$) random mixtures of uniform
spheres with effective volume concentrations $f \in [0,1]$; \item
computer simulations \cite{Karkainen2001} for
intermediate-contrast ($\varepsilon_1/\varepsilon_0 = 51$)
dispersions of hard  ($c \in [0,0.3]$) and freely overlapping ($f
\in [0,1]$) uniform spheres.
\end{enumerate} The agreement of the results  with calculations \cite{Felderhof1982a,Torquato1984,Torquato1985,
Jeffrey1973,Sihvola1989} is very good and with equation
\cite{Jylha2007} satisfactorily good. The agreement with
simulation data \cite{Karkainen2001} is achieved under the
assumption that an uncontrolled use of weighted averages for the
local permittivity values near the surfaces of spheres, which is
typical of the finite difference method, is equivalent to the
introduction of an interphase layer (of certain thickness and
permittivity) between the spheres and the matrix. The neglect of
this fact may cause considerable computational errors when
rectangular grids are used to simulate dielectric properties of
systems of particles with arbitrary oriented surfaces.
\end{enumerate}

\end{document}